\documentclass[12pt]{article}

\usepackage{amsmath,amsfonts,amssymb}

\begin{document}

\begin{titlepage}

\begin{center}

\begin{center}
{\Large{ \bf Integrability of D1-brane on Group Manifold  with Mixed
Three Form Flux}}
\end{center}

\vskip 1cm

{\large Josef Kluso\v{n}$^{}$\footnote{E-mail: {\tt
klu@physics.muni.cz}} }

\vskip 0.8cm

{\it Department of
Theoretical Physics and Astrophysics\\
Faculty of Science, Masaryk University\\
Kotl\'{a}\v{r}sk\'{a} 2, 611 37, Brno\\
Czech Republic\\
[10mm]}

\vskip 0.8cm

\end{center}

\begin{abstract}
We consider D1-brane as a natural probe of the group manifold with
mixed three form fluxes.  We determine Lax connection for given
theory. Then we switch to the canonical analysis and calculate the
Poisson brackets between spatial components of Lax connections and
we argue for integrability of given theory.

\end{abstract}

\bigskip

\end{titlepage}

\newpage

\newcommand{\mT}{\mathcal{T}}
\def\tr{\mathrm{Tr}}
\def\tJ{\tilde{J}}
\def\str{\mathrm{Str}}
\newcommand{\tL}{\tilde{L}}
\def\bV{\mathbf{V}}
\def\Pf{\mathrm{Pf}}
\def\ttheta{\tilde{\theta}}
\def\bX{\mathbf{X}}
\def\bY{\mathbf{Y}}
\def\I{\mathbf{i}}
\def\IT{\I_{\Phi,\Phi',T}}
\def\tPi{\tilde{\Pi}}
\def\tmH{\tilde{\mH}}
\def\bC{\mathbf{C}}
\def \cit{\IT^{\dag}}
\def\hPi{\hat{\Pi}}
\def \cdt{\overline{\tilde{D}T}}
\def \dt{\tilde{D}T}
\def\bra #1{\left<#1\right|}
\def\ket #1{\left|#1\right>}
\def\vac #1{\left<\left<#1\right>\right>}
\def\pb  #1{\left\{#1\right\}}
\def \uw #1{(w^{#1})}
\def\bX{\mathbf{X}}
\def\bB{\mathbf{B}}
\def\bC{\mathbf{C}}
\def \dw #1{(w_{#1})}
\newcommand{\bK}{\mathbf{K}}
\newcommand{\thw}{\tilde{\hat{w}}}
\newcommand{\bA}{{\bf A}}
\newcommand{\bd}{{\bf d}}
\newcommand{\bD}{{\bf D}}
\newcommand{\bF}{{\bf F}}
\newcommand{\bN}{{\bf N}}
\newcommand{\hp}{\hat{p}}
\newcommand{\hq}{\hat{q}}
\newcommand{\hF}{\hat{F}}
\newcommand{\hG}{\hat{G}}
\newcommand{\hH}{\hat{H}}
\newcommand{\hU}{\hat{U}}
\newcommand{\mH}{\mathcal{H}}
\newcommand{\mG}{\mathcal{G}}
\newcommand{\mA}{\mathcal{A}}
\newcommand{\mD}{\mathcal{D}}
\newcommand{\tpr}{t^{\prime}}
\newcommand{\bzg}{\overline{\zg}}
\newcommand{\of}{\overline{f}}
\newcommand{\ow}{\overline{w}}
\newcommand{\htheta}{\hat{\theta}}
\newcommand{\opartial}{\overline{\partial}}
\newcommand{\hd}{\hat{d}}
\def\mF{\mathcal{F}}
\newcommand{\halpha}{\hat{\alpha}}
\newcommand{\hbeta}{\hat{\beta}}
\newcommand{\hdelta}{\hat{\delta}}
\newcommand{\hgamma}{\hat{\gamma}}
\newcommand{\hlambda}{\hat{\lambda}}
\newcommand{\hw}{\hat{w}}
\newcommand{\hN}{\hat{N}}
\newcommand{\onabla}{\overline{\nabla}}
\newcommand{\hmu}{\hat{\mu}}
\newcommand{\hnu}{\hat{\nu}}
\newcommand{\ha}{\hat{a}}
\newcommand{\hb}{\hat{b}}
\newcommand{\hc}{\hat{c}}
\newcommand{\com}[1]{\left[#1\right]}
\newcommand{\oz}{\overline{z}}
\newcommand{\oJ}{\overline{J}}
\newcommand{\mL}{\mathcal{L}}
\newcommand{\oh}{\overline{h}}
\newcommand{\oT}{\overline{T}}
\newcommand{\oepsilon}{\overline{\epsilon}}
\newcommand{\tP}{\tilde{P}}
\newcommand{\hP}{\hat{P}}
\newcommand{\talpha}{\tilde{\alpha}}
\newcommand{\uc}{\underline{c}}
\newcommand{\ud}{\underline{d}}
\newcommand{\ue}{\underline{e}}
\newcommand{\uf}{\underline{f}}
\newcommand{\hpi}{\hat{\pi}}
\newcommand{\oZ}{\overline{Z}}
\newcommand{\tg}{\tilde{g}}
\newcommand{\tK}{\tilde{K}}
\newcommand{\tj}{\tilde{j}}
\newcommand{\tG}{\tilde{G}}
\newcommand{\hg}{\hat{g}}
\newcommand{\htG}{\hat{\tilde{G}}}
\newcommand{\hX}{\hat{X}}
\newcommand{\hY}{\hat{Y}}
\newcommand{\bDi}{\left(\bD^{-1}\right)}
\newcommand{\hthteta}{\hat{\theta}}
\newcommand{\hB}{\hat{B}}
\newcommand{\tlambda}{\tilde{\lambda}}
\newcommand{\thlambda}{\tilde{\hat{\lambda}}}
\newcommand{\tw}{\tilde{w}}
\newcommand{\hJ}{\hat{J}}
\newcommand{\tPsi}{\tilde{\Psi}}
\newcommand{\cP}{{\cal P}}
\def\ba{\mathbf{a}}
\newcommand{\tphi}{\tilde{\phi}}
\newcommand{\tOmega}{\tilde{\Omega}}
\newcommand{\homega}{\hat{\omega}}
\newcommand{\hupsilon}{\hat{\upsilon}}
\newcommand{\hUpsilon}{\hat{\Upsilon}}
\newcommand{\hOmega}{\hat{\Omega}}
\newcommand{\bJ}{\mathbf{J}}
\newcommand{\olambda}{\overline{\lambda}}
\newcommand{\uhlambda}{\underline{\hlambda}}
\newcommand{\uhw}{\underline{\hw}}
\newcommand{\tC}{\tilde{C}}
\def \lhw #1{(\hw^{#1})}
\def \dhw #1{(\hw_{#1})}
\newcommand{\bG}{\mathbf{G}}
\newcommand{\bhG}{\hat{\bG}}
\newcommand{\bH}{\mathbf{H}}
\newcommand{\bE}{\mathbf{E}}
\newcommand{\mJ}{\mathcal{J}}
\newcommand{\mY}{\mathcal{Y}}
\newcommand{\mZ}{\mathcal{Z}}
\newcommand{\hj}{\hat{j}}
\newcommand{\bAi}{\left(\bA^{-1}\right)}
\newcommand{\hpartial}{\hat{\partial}}
\newcommand{\hD}{\hat{D}}
\newcommand{\mC}{\mathcal{C}}
\newcommand{\omC}{\overline{\mC}}
\newcommand{\mP}{\mathcal{P}}
\newcommand{\omP}{\overline{\mP}}
\newcommand{\tLambda}{\tilde{\Lambda}}
\newcommand{\tPhi}{\tilde{\Phi}}
\newcommand{\tD}{\tilde{D}}
\newcommand{\tgamma}{\tilde{\gamma}}
\section{Introduction and Summary }
The integrability in the AdS/CFT correspondence is fundamental for
the calculations beyond perturbative theory. Famous example is the
duality between $\mathcal{N}=4$ super-Yang-Mills theory in four
dimensions and Type IIB theory on $AdS_5\times S^5$ with the
Ramond-Ramond (RR) flux, where the exact string spectrum and the
spectrum of anomalous dimensions in the SYM theory can be described
by Bethe-ansatz equations \footnote{For review, see
\cite{Sfondrini:2014via,Puletti:2010ge,vanTongeren:2013gva,Beisert:2010jr}.}.
The integrability on the string theory side of the correspondence is
based on the existence of the Lax connection that implies an
existence of infinite number of conserved charges
\cite{Bena:2003wd}. However this is only necessary  condition since
the integrability of the theory also requires  that these conserved
charges are in involution  as was stressed in \cite{Dorey:2006mx}.

It is well known that integrability can be applied for  group
manifolds with non-trivial RR and NSNS fluxes. Such a famous example
is string theory on $AdS_3$ background with non-trivial RR and NSNS
fluxes. It turns out that in case of  pure NSNS flux the string
theory can be quantized using world-sheet conformal field theory
techniques \cite{Giveon:1998ns,Elitzur:1998mm,
deBoer:1998pp,Maldacena:2000hw,Maldacena:2000kv,Maldacena:2001km}.
On the other hand the RR $AdS_3$ backgrounds have more complicated
CFT description \cite{Berkovits:1999im} while these backgrounds are
integrable as well \cite{Chen:2005uj,Babichenko:2009dk}.

On the other hand the case of mixed RR/NSNS $AdS_3$ background is
much more challenging either from CFT perspective or from the
integrability point of view. One possibility is to consider small
derivations from the pure NSNS point using conformal perturbative
theory \cite{Quella:2007sg}. Another possibility was suggested in
\cite{Cagnazzo:2012se}, where the starting point was pure RR
background  with new WZ term that represents the coupling to the
NSNS flux. This beautiful construction  leads to the rapid progress in
understanding of the role integrability in theory with mixed fluxes,
for related works, see
\cite{Hoare:2013pma,Hoare:2013ida,Hoare:2013lja,
Babichenko:2014yaa,Lloyd:2014bsa,Sundin:2014ema}.

It is well known that the $AdS_3$ backgrounds with different  fluxes
are related by U-duality transformations. For example, type IIB
S-duality relates $AdS_3\times S^3\times T^4$ backgrounds supported
by different three-form fluxes: the pure RR flux background arises
as the near horizon limit of D1-D5-brane system while background
supported by mixed three form flux involves the near horizon limit
of NS5-branes and fundamental strings in addition to the D1 and
D5-branes. At the same time fundamental string transforms under
S-duality to the bound state of D1-brane and fundamental string.
Then one can ask the question whether D1-brane could be considered
as another probe  in string theory that naturally incorporates the
coupling between NSNS and RR fields. In fact, the low energy
descriptions of D1-brane  is given by Dirac-Born-Infeld action
together with Chern-Simons term with explicit coupling to RR and
NSNS two forms. We demonstrated in our previous paper
\cite{Kluson:2014uaa} that D1-brane on the group manifold with
non-trivial NSNS flux is integrable. In this paper we extend given
analysis to the most general background including dilaton,
Ramond-Ramond zero form $C^{(0)}$ and Ramond-Ramond two form
$C^{(2)}$ together with the three forms $F=dC^{(2)}, H=dB$ that can
be expressed using the structure constants of the group that defines
the group manifold on which D1-brane propagates. We find that this
D1-brane is integrable on condition when dilaton and Ramond-Ramond
zero form are constants. Then we perform canonical analysis of given
theory and calculate the Poisson brackets between spatial components
of Lax connections. We show that this Poisson bracket has the form
that ensures that the conserved charges are in involutions up to the
well known problems with terms containing derivative of delta
functions that need  special regularizations
\cite{Maillet:1985ek,Delduc:2012qb,Delduc:2012vq}. Then we consider
concrete example which is D1-brane on $AdS_3\times S^3$ background
with mixed RR/NS flux. We firstly show that the equation of motion
for this D1-brane can be expressed as the equation of conservation
of specific current which is however non-linear due to the specific
form of D-brane action. Then introducing an auxiliary metric and
corresponding constraint we can rewrite this current to the
manifestly linear form \footnote{This is  similar situation as in
the case of the equivalence between Nambu-Gotto and Polyakov action
for bosonic string.}. Then fixing the gauge and for certain
backgrounds we can find currents whose conservation law is special
and that is an analogue of the holomorphic and anti-holomorphic
currents in Wess-Zumino-Witten model \cite{Witten:1983ar}.
Explicitly, we find that this occurs in case of D1-brane in the near
horizon limit of D1-D5-brane background with zero electric flux.
Surprisingly we also find that the same situation occurs in case of
the background with non-zero RR and NSNS fluxes that arises from
D1-D5-brane background through specific $SL(2,Z)$ transformation.
 This is very
interesting result that suggests the possibility that for these
values of fluxes D1-brane theory can be treated with the help of
powerful techniques of two dimensional conformal field theory.

Let us outline our results. We show that  D1-brane can be considered
as a natural probe of backgrounds with mixed flux. We mean that
given idea is very attractive and should be elaborated further. For
example, it would be nice to explicitly determine world-sheet
S-matrix for given theory in the $AdS_3$ background with mixed flux.
It would be also nice to analyze classical solutions on the
world-volume of given theory corresponding to possible magnon
solutions and compare them with the string solutions. We hope to
return to these problems in future. It would be also interesting try
to extend given analysis to the supersymmetric  D1-brane theory.
Further question that deserves detailed treatment is the question of
the conformal field theory description of D1-brane with electric
flux on $AdS_3\times S^3$ with specific values of fluxes. We hope to
return to all these problems in future.

This paper is organized as follows. In the next section
(\ref{second})  we introduce D1-brane on the group manifold
background with non-trivial NSNS  and two RR forms. We analyze under
which conditions is the world-sheet theory integrable. Then in
section (\ref{third}) we perform Hamiltonian analysis of given
theory and calculate the Poisson brackets between spatial components
of Lax connection. Finally in section (\ref{fourth}) we consider
D1-brane on various  $AdS_3\times S^3$ backgrounds with three form
fluxes.

\section{D1-brane on Group Manifold}
\label{second} In this section we introduce D1-brane action that
governs the dynamics of D1-brane on general background. Recall that
given action is the sum of DBI and CS term and has the form
\begin{eqnarray}\label{D1branegen}
S&=&-T_{D1}\int d\tau d\sigma e^{-\Phi}\sqrt{-\det
\bA}+\nonumber \\
&+&T_{D1}\int d\tau d\sigma
((b_{\tau\sigma}+2\pi\alpha'\mF_{\tau\sigma})C^{(0)}+
c_{\tau\sigma}) \ ,
\nonumber \\
\bA_{\alpha\beta}&=&G_{MN}\partial_\alpha x^M
\partial_\beta x^N+2\pi\alpha'\mF_{\alpha\beta}+B_{MN}
\partial_\alpha x^M\partial_\beta x^N \ , \nonumber \\
\mF_{\alpha\beta}&=&\partial_\alpha A_\beta-\partial_\beta A_\alpha
\
, \nonumber \\
\end{eqnarray}
where $x^M,M,N=0,1,\dots,D$ are embedding coordinates of D1-brane in
the background that is specified by the metric $G_{MN}(X)$ and NSNS
two form  $B_{MN}=-B_{NM}$ together with Ramond-Ramond two form
$C^{(2)}_{MN}=-C^{(2)}_{NM}$. We further consider background with
non-trivial dilaton $\Phi$ and RR zero form $C^{(0)}$. Further,
$\sigma^\alpha=(\tau,\sigma)$ are world-sheet coordinates of
D1-brane and $b_{\tau\sigma},c_{\tau\sigma}$ are pull-backs of
$B_{MN}$ and $C_{MN}$ to the world-volume of D1-brane. Explicitly,
\begin{equation}
b_{\alpha\beta}\equiv B_{MN}\partial_\alpha x^M\partial_\beta x^N=
-b_{\beta\alpha} \ , \quad  c_{\tau\sigma}=C^{(2)}_{MN}\partial_\tau
x^M\partial_\sigma x^N \ .
\end{equation}
Finally $T_{D1}=\frac{1}{2\pi\alpha'}$ is D1-brane tension and
$A_{\alpha},\alpha=\tau,\sigma$ is two dimensional gauge field that
propagates on the world-sheet of D1-brane.

 It is useful to
rewrite the action (\ref{D1branegen}) into the form
\begin{eqnarray}\label{D1actional}
S&=&-T_{D1}\int d\tau d\sigma e^{-\Phi} \sqrt{-\det g-
(2\pi\alpha'\mF_{\tau\sigma}+b_{\tau\sigma})^2}+ \nonumber \\
&+&T_{D1}\int d\tau d\sigma
((b_{\tau\sigma}+2\pi\alpha'\mF_{\tau\sigma})C^{(0)}+
c_{\tau\sigma})
\ , \nonumber \\
\end{eqnarray}
where $g_{\alpha\beta}=G_{MN}\partial_\alpha x^M
\partial_\beta x^N, \det g=g_{\tau\tau}g_{\sigma\sigma}-
(g_{\tau\sigma})^2$. From (\ref{D1actional}) we obtain the equations
of motion for $x^M$
\begin{eqnarray}\label{eqxM}
& &\partial_M [\Phi]e^{-\Phi}\sqrt{-\det g-
(2\pi\alpha'\mF_{\tau\sigma}+b_{\tau\sigma})^2}-\nonumber \\
&-&\partial_\alpha\left[\frac{G_{MN}\partial_\beta x^N
g^{\beta\alpha} \det g} {\sqrt{-\det g-
(2\pi\alpha'\mF_{\tau\sigma}+b_{\tau\sigma})^2}}\right]
+\frac{\partial_M G_{KL}\partial_\alpha x^K \partial_\beta x^L
g^{\beta\alpha}\det g}{2 {\sqrt{-\det g-
(2\pi\alpha'\mF_{\tau\sigma}+b_{\tau\sigma})^2}}}
+ \nonumber \\
&+&\frac{(2\pi\alpha'\mF_{\tau\sigma}+b_{\tau\sigma})} {\sqrt{-\det
g-(2\pi\alpha'\mF_{\tau\sigma}+b_{\tau\sigma})^2}}
\partial_M B_{KL}\partial_\tau x^ K\partial_\sigma x^L
-\nonumber \\
&-&\partial_\tau \left[ \frac{B_{MN}\partial_\sigma x^N
(2\pi\alpha'\mF_{\tau\sigma}+b_{\tau\sigma})} {\sqrt{-\det
g-(2\pi\alpha'\mF_{\tau\sigma}+b_{\tau\sigma})^2}}\right]
+\partial_\sigma \left[ \frac{B_{MN}\partial_\tau x^N
(2\pi\alpha'\mF_{\tau\sigma}+b_{\tau\sigma})} {\sqrt{-\det
g-(2\pi\alpha'\mF_{\tau\sigma}+b_{\tau\sigma})^2}}
\right]+ \nonumber \\
&+&\partial_M C^{(0)}(b_{\tau\sigma}+2\pi\alpha'\mF_{\tau\sigma})+
C^{(0)}\partial_M b_{KL}\partial_\tau x^K\partial_\sigma x^L-\nonumber \\
&-&\partial_\tau[C^{(0)}b_{MK}\partial_\sigma x^K]-
\partial_\sigma[C^{(0)}b_{KM}\partial_\tau x^K]+\nonumber \\
&+&\partial_M C^{(2)}_{KL}\partial_\tau x^K\partial_\sigma
x^L-\partial_\tau [C_{MK}^{(2)}\partial_\sigma x^K]
-\partial_\sigma[C_{KM}^{(2)}\partial_\tau x^K]
 =0 \nonumber \\
\end{eqnarray}
while the equations of motion for $A_\tau,A_\sigma$ take the form
\begin{eqnarray}
\partial_\tau\left[e^{-\Phi}\frac{(2\pi\alpha'\mF_{\tau\sigma}+b_{\tau\sigma})}
{\sqrt{-\det g-
(2\pi\alpha'\mF_{\tau\sigma}+b_{\tau\sigma})^2}}+C^{(0)}\right]=0 \
,
\nonumber \\
\partial_\sigma\left[e^{-\Phi}\frac{(2\pi\alpha'\mF_{\tau\sigma}+b_{\tau\sigma})}
{\sqrt{-\det g-
(2\pi\alpha'\mF_{\tau\sigma}+b_{\tau\sigma})^2}}+C^{(0)}\right]=0 \
.
\nonumber \\
\end{eqnarray}
Last two equations imply an existence of constant electric flux
\begin{equation}
\frac{e^{-\Phi}(2\pi\alpha'\mF_{\tau\sigma}+b_{\tau\sigma})}
{\sqrt{-\det
g-(2\pi\alpha'\mF_{\tau\sigma}+b_{\tau\sigma})^2}}+C^{(0)} =\Pi \ ,
\quad \Pi=\mathrm{const} \ .
\end{equation}
With the help of this constant we can express
$2\pi\alpha'\mF_{\tau\sigma}+b_{\tau\sigma}$ as
\begin{eqnarray}\label{FplusB}
2\pi\alpha'\mF_{\tau\sigma}+b_{\tau\sigma}= \frac{(\Pi-C^{(0)})
\sqrt{-\det
g}}{\sqrt{e^{-2\Phi}+(\Pi-C^{(0)})^2}} \ , \nonumber \\
\end{eqnarray}
so that the equations of motion (\ref{eqxM}) simplify considerably
\begin{eqnarray}\label{eqxM2}
&-&\partial_M[\sqrt{e^{-2\Phi}+(\Pi-C^{(0)})^2}]\sqrt{-\det g}+ \nonumber \\
&+&\partial_\alpha\left[G_{MN}\partial_\beta x^N g^{\beta\alpha}
\sqrt{-\det g}\sqrt{e^{-2\Phi}+(\Pi-C^{(0)})^2}\right]
-\nonumber \\
&-&\frac{1}{2}\partial_M G_{KL}\partial_\alpha x^K \partial_\beta
x^L \sqrt{-\det g}\sqrt{ e^{-2\Phi}+(\Pi-
C^{(0)})^2}+\nonumber \\
&+& \Pi H_{MKN}\partial_\tau x^K\partial_\sigma x^N+F_{MKN}
\partial_\tau x^K\partial_\sigma x^N=0 \ ,
\nonumber \\
\end{eqnarray}
where
\begin{eqnarray}
H_{MNK}&=&\partial_M B_{NK}+\partial_N B_{KM}+
\partial_K B_{MN} \ , \nonumber \\
F_{MNK}&=&\partial_M C^{(2)}_{NK}+\partial_N C^{(2)}_{KM}+
\partial_K C^{(2)}_{MN} \ .
\nonumber \\
\end{eqnarray}
Now we are going to be more specific about the background. When we
consider group manifold $G$ we presume that
 the metric $G_{MN}$ can be expressed as
\begin{equation}
G_{MN}=E_M^{ \ A}E_N^{ \ B}K_{AB} \ ,
\end{equation}
where for the group element $g\in G$ we have
\begin{equation}\label{defEM}
J\equiv g^{-1}dg=E_M^{ \  A}T_A dx^M \ ,
\end{equation}
where $T_A$ is the basis of Lie Algebra $\mG$ of the group $G$. Note
that $K_{AB}=\tr (T_A T_B)$. Further, from the definition
(\ref{defEM}) we obtain
\begin{equation}
dJ+J\wedge J=0
\end{equation}
that implies an important relation
\begin{equation}\label{relE}
\partial_M E_N^{ \ A}-\partial_N E_M^{ \ A}+f^A_{ \ BC}E_M^{  \
B}E_N^{  \ C}=0 \ ,
\end{equation}
where
\begin{equation}
[T_B,T_C]=T_A f^A_{ \ BC} \ .
\end{equation}
In case of the   fluxes $F_{KLM},H_{KLM}$ we presume following
relations between them and the  structure constants $f_{ABC}$ of the
Lie algebra $\mathcal{G}$
\begin{equation}\label{HFdef}
H_{MNK}E^{M}_{ \ A}E^N_{ \ B} E^K_{ \ C}=\kappa f_{ABC} \ , \quad
F_{MNK}E^{M}_{ \ A}E^N_{ \ B} E^K_{ \ C}= \omega f_{ABC} \ ,
\end{equation}
where $\kappa$ and $\omega$ are constants. The first one formula is
well known relation that defines Wess-Zumino term when we describe
motion of string on group space with $B-$flux \footnote{See for
example \cite{Evans:2000hx} for nice discussion and calculations of
Poisson brackets of various currents.}.  In case of Ramond-Ramond
flux we introduce this relation in order to preserve symmetry
between NS-NS and RR fluxes. At this place we will not discuss  the
problem whether background fields define a consistent string theory
background and hence we can consider $\kappa$ and $\omega$ as free
parameters. On the other hand it is important to stress that when we
discuss D1-brane on $AdS_3\times S^3$ with mixed fluxes these
coefficients $\kappa$ and $\omega$ have concrete values in order to
define consistent string theory background. We will discuss this
case in more details in section (\ref{fourth}).

With the help of (\ref{HFdef})  we can write
\begin{eqnarray}\label{EHx}
E^M_{ \ C}H_{MKL}\partial_\tau x^K \partial_\sigma x^L
&=& \kappa  f_{CAB}J^A_\tau J^B_\sigma \nonumber \\
E^M_{ \ C}F_{MKL}\partial_\tau x^K \partial_\sigma x^L&=& \omega
f_{CAB}J^A_\tau J^B_\sigma \ .
\end{eqnarray}
Note that $E^M_{ \ A}$ is inverse to $E_M^{\ B}$ defined as
\begin{equation}
E^M_{ \ A}E_M^{ \ B}=\delta_A^B \ , \quad E^M_{  \ A}E_N^{ \ A}=
\delta^M_N \ .
\end{equation}
 Now with the help of
(\ref{relE}) and (\ref{EHx}) we can rewrite the equations of motion
(\ref{eqxM2}) to the form that contains the current
$J_\alpha^A=E_M^{ \ A}\partial_\alpha x^M$
\begin{eqnarray}\label{eqfinal}
&-&E^M_{ \ C}\partial_M[\sqrt{e^{-2\Phi}+(\Pi-C^{(0)})^2}]\sqrt{-\det g}+ \nonumber \\
 &+&K_{CB}\partial_\alpha[J_\beta^B g^{\beta\alpha}
\sqrt{-\det g}\sqrt{e^{-2\Phi}+(\Pi-
C^{(0)})^2}]+\nonumber \\
 &+&\Pi \kappa f_{CAB}J^A_\tau J^B_\sigma+
 \omega f_{CAB}J^A_\tau J^B_\sigma =0 \ .  \nonumber \\
\end{eqnarray}
Now we are ready to analyze the integrability of given theory. Let
us consider following current
\begin{eqnarray}\label{LA}
L_\tau^A=AJ_\tau^A+B\sqrt{-g}g^{\sigma\alpha}\sqrt{e^{-2\Phi}+(\Pi-C^{(0)})^2}J_\alpha^A
\ ,
\nonumber \\
L_\sigma^A=AJ_\sigma^A-B\sqrt{-g}g^{\tau\alpha}\sqrt{e^{-2\Phi}+(\Pi-C^{(0)})^2}J_\alpha^A
\ ,
\nonumber \\
\end{eqnarray}
where $A$ and $B$ are coefficients that will be determined by
requirement that the current $L^A_\alpha$ is  flat. First of all we
calculate
\begin{eqnarray}
\partial_\tau L_\sigma^A-\partial_\sigma L_\tau^A
&=&-AJ_\tau^BJ_\sigma^Cf_{\ BC}^A+B(\Pi\kappa+\omega)
f^A_{\ BC}J^B_\tau J^C_\sigma- \nonumber \\
&-&K^{AB}E^M_{ \
B}\partial_M[\sqrt{e^{-2\Phi}+(\Pi-C^{(0)})^2}]\sqrt{-\det g} \ ,
\nonumber \\
\end{eqnarray}
where we used the equations of motion (\ref{eqfinal}) together with
the condition (\ref{relE}). As the next step we calculate
\begin{eqnarray}
f^A_{\ BC}L^B_\tau L^C_\sigma=
\left(A^2-B^2\left[e^{-2\Phi}+(\Pi-C^{(0)})^2\right]\right)
 f^A_{\ BC}J^B_\tau J^C_\sigma \ .  \nonumber \\
\end{eqnarray}
Collecting these two results together we obtain
\begin{eqnarray}\label{Lflat}
& &\partial_\tau L_\sigma^A-\partial_\sigma
L_\tau^A+f^A_{BC}L^B_\tau L^C_\sigma=\nonumber
\\
&=&(-A+B(\Pi\kappa+\omega)+
A^2-B^2\left[e^{-2\Phi}+(\Pi-C^{(0)})^2\right])f^A_{ \ BC}J^B_\tau
J^C_\sigma-\nonumber \\
&-& K^{AB}E^M_{ \
B}\partial_M[\sqrt{e^{-2\Phi}+(\Pi-C^{(0)})^2}]\sqrt{-\det g} \ .
\nonumber \\
\end{eqnarray}
Let us now discuss the result derived above. The expression on the
second line is proportional to the currents while the expression on
the third line contains derivatives of the background fields
$C^{(0)}$ and $\Phi$. Then clearly expressions on the second and
third line have to vanish separately in order  $L^A_\alpha$ to be
flat. The expression on the third line vanishes when we require that
 $\sqrt{e^{-2\Phi}+(\Pi-C^{(0)})^2}$ is
constant.
This can be ensured for non-zero electric flux $\Pi$ when $\Phi$ and
$C^{(0)}$ are constant. Then we have to demand that the expression
on the second line in (\ref{Lflat}) vanishes. If we consider the
ansatz $B=-\Lambda A$ we find the solutions in the form
\begin{eqnarray}\label{ABsol}
A&=&\frac{1}{1-\Lambda^2(e^{-2\Phi}+(\Pi-C^{(0)})^2)}(1+(\Pi
\kappa +\omega)\Lambda) \ , \nonumber \\
 B&=&-\frac{\Lambda}{
1-\Lambda^2(e^{-2\Phi}+(\Pi-C^{(0)})^2)}(1+(\Pi \kappa
+\omega)\Lambda) \ ,
\nonumber \\
\end{eqnarray}
where $\Lambda$ is a spectral parameter.  Finally we should mention
that this is on-shell condition. On the other hand if we calculate
the Poisson bracket between these currents we have to express $A$
and $B$ given in (\ref{ABsol})using off-shell form of the
combinations $e^{-2\Phi}+(\Pi-C^{(0)})^2$ and $\Pi$. Explicitly,
from (\ref{FplusB}) we obtain
\begin{eqnarray}\label{onoffshellrelation}
& &e^{-2\Phi}+(\Pi-C^{(0)})^2=\frac{e^{-2\Phi} \det g}{ \det
g+(2\pi\alpha'\mF_{\tau\sigma}+b_{\tau\sigma})^2} \ , \nonumber \\
& &\Pi= \frac{e^{-\Phi}(2\pi\alpha'\mF_{\tau\sigma}+b_{\tau\sigma})}
{\sqrt{-\det
g-(2\pi\alpha'\mF_{\tau\sigma}+b_{\tau\sigma})^2}}+C^{(0)} \ .
\nonumber \\
\end{eqnarray}
Inserting (\ref{ABsol}) and (\ref{onoffshellrelation}) into
(\ref{LA}) we find off-shell formulation of flat current. In the
next section we express the spatial components of the flat current
using canonical variables and calculate Poisson bracket between
them.

Let us summarize results derived in this section. We studied the
dynamics of D1-brane on the group manifold with non-trivial NSNS and
RR two form fluxes and together with dilaton and RR zero form. We
argued that it is possible to define Lax connection for this theory
and we showed that this Lax connection is flat on condition when the
dilaton and RR zero form are constant. The existence of the Lax
connection is a necessary condition of integrability. The additional
condition is that corresponding conserved charges are in involution
which can be seen from the form of the Poisson bracket between
spatial components of Lax connection. The calculation of this
Poisson bracket will be performed in the next section.

\section{Poisson Brackets of Lax Connection}\label{third}
In this section we  calculate the Poisson brackets between spatial
components of  Lax connection. To do this we have to develop the
Hamiltonian formalism for $D1-$brane action in general background.
We start with the action (\ref{D1actional}) and find corresponding
conjugate momenta
\begin{eqnarray}\label{defpM}
& &p_M=\frac{\delta L}{\delta \partial_\tau x^M}= T_{D1}
\frac{e^{-\Phi}}{\sqrt{-\det g -(2\pi\alpha'F_{\tau\sigma}+
b_{\tau\sigma})^2}}(G_{MN}\partial_\alpha x^N g^{\alpha \tau}\det g+\nonumber \\
& &+(2\pi\alpha'F_{\tau\sigma}+ b_{\tau\sigma})B_{MN}\partial_\sigma
x^N)+T_{D1}(C^{(0)}B_{MN}\partial_\sigma x^N+C^{(2)}_{MN}\partial_\sigma x^N) \ , \nonumber \\
& &\pi^\sigma=\frac{\delta L}{\delta
\partial_\tau A_\sigma}=\frac{e^{-\Phi}(2\pi\alpha'F_{\tau\sigma}+
b_{\tau\sigma})}{\sqrt{-\det g -(2\pi\alpha'F_{\tau\sigma}+
b_{\tau\sigma})^2}}+C^{(0)}\ , \quad \pi^\tau=\frac{\delta L}{\delta
\partial_\tau A_
\tau}\approx 0 \nonumber \\
\end{eqnarray}
and hence
\begin{eqnarray}
\Pi_M&\equiv&
p_M-\frac{\pi^\sigma}{(2\pi\alpha')}B_{MN}\partial_\sigma x^N
-T_{D1}(C^{(0)}B_{MN}\partial_\sigma x^N+C^{(2)}_{MN}\partial_\sigma
x^N)= \nonumber \\
&=&T_{D1} \frac{e^{-\Phi}}{\sqrt{-\det g
-(2\pi\alpha'F_{\tau\sigma}+
b_{\tau\sigma})^2}}G_{MN}\partial_\alpha x^N g^{\alpha \tau}\det g \
.
\nonumber \\
\end{eqnarray}
Using these relations it is easy to see that the bare Hamiltonian is
equal to
\begin{eqnarray}
H_B=\int d\sigma(p_M\partial_\tau x^M+\pi^\sigma \partial_\tau
A_\sigma-\mL)= \int d\sigma \pi^\sigma\partial_\sigma A_\tau
\nonumber \\
\end{eqnarray}
while we have three primary constraints
\begin{eqnarray}
& &\pi^\tau\approx 0 \ , \quad \mH_\sigma\equiv p_M\partial_\sigma x^M\approx 0 \ , \nonumber \\
& &\mH_\tau\equiv\frac{1}{T_{D1}} \Pi_M G^{MN}\Pi_N+
T_{D1}\left(e^{-2\Phi}+\left(\pi^\sigma -C^{(0)}\right)^2
\right)g_{\sigma\sigma}\approx 0
\ . \nonumber \\
\end{eqnarray}
  Including
these primary constraints to the definition of the Hamiltonian we
obtain an extended Hamiltonian in the form
\begin{equation}
H=\int d\sigma (\lambda_\tau\mH_\tau+\lambda_\sigma
\mH_\sigma-A_\tau\partial_\sigma\pi^\sigma+v_\tau \pi^\tau) \ ,
\end{equation}
where $\lambda_\tau,\lambda_\sigma,v_\tau$ are Lagrange multipliers
corresponding to the primary constraints $\mH_\tau\approx 0 \
,\mH_\sigma\approx 0 \ , \pi^\tau\approx 0$.
Now we have to check the stability of all constraints. The
requirement of the preservation of the primary constraint
$\pi^\tau\approx 0$ implies the secondary constraint
\begin{equation}
\mG=\partial_\sigma \pi^\sigma\approx 0 \ .
\end{equation}
In case of  the constraints $\mH_\tau,\mH_\sigma$ we can easily show
in the same way as in \cite{Kluson:2014uaa} that the constraints
$\mH_\tau,\mH_\sigma$ are first class constraints and hence they
preserved during the time evolution.

Now we are ready to proceed to the calculations of the Poisson
brackets between spatial components of the flat current $L_\sigma^A$
for different spectral parameters $\Lambda$ and $\Gamma$
\begin{eqnarray}\label{pblAB}
\pb{L^A_\sigma(\Lambda,\sigma),L^B_\sigma(\Gamma,\sigma')} \ .
\nonumber
\\
\end{eqnarray}
 Recall that these are currents that define monodromy matrix and
hence corresponding conserved charges.
Using (\ref{onoffshellrelation}) and (\ref{defpM}) we find that the
spatial component of the current $L^A_\sigma$  expressed using
canonical variables has the form
\begin{equation}
L^A_\sigma=\frac{1+\Lambda(\pi^\sigma\kappa+\omega)}{1-\Lambda^2(e^{-2\Phi}+(\pi^\sigma-C^{(0)})^2)}
\left(E_M^{ \ A}\partial_\sigma x^M-\frac{\Lambda}{T_{D1}}E^M_{ \
B}K^{AB}\Pi_M\right) \ .
\end{equation}
In order to calculate  (\ref{pblAB}) we need following  Poisson
brackets
\begin{eqnarray}
\pb{x^M(\sigma),\Pi_N(\sigma')}&=&\delta^M_N\delta(\sigma-\sigma') \
,
\nonumber \\
\pb{E_M^{ \ A}(\sigma),\Pi_N(\sigma')}&=&\partial_N E_M^{ \
A}\delta(\sigma-\sigma') \ , \nonumber \\
\pb{E^M_{ \ A}(\sigma),\Pi_N(\sigma')}&=&\partial_N E^M_{ \ A}
\delta(\sigma-\sigma') \ , \nonumber \\
\end{eqnarray}
and also
\begin{eqnarray}
& &\pb{\Pi_M(\sigma),\Pi_N(\sigma')}=\nonumber \\
&=& \frac{1}{2\pi\alpha'} (\pi^\sigma+C^{(0)})H_{MNK}\partial_\sigma
x^K\delta(\sigma-\sigma')+\frac{1}{2\pi\alpha'}F_{MNK}\partial_\sigma
x^K \delta(\sigma-\sigma')+
\frac{1}{2\pi\alpha'}\mG B_{MN}\delta(\sigma-\sigma')\nonumber \\
\end{eqnarray}
and finally
\begin{eqnarray}
& &\pb{E^M_{ \ A}\Pi_M(\sigma),E^N_{\ B}\Pi_N(\sigma')}=
-E^M_{ \ D}f^D_{ \ AB}\Pi_M\delta(\sigma-\sigma')  + \nonumber \\
&+& E^M_{ \ A}\left(\frac{1}{2\pi\alpha'}
(\pi^\sigma+C^{(0)})H_{MNK}\partial_\sigma
x^K+\frac{1}{2\pi\alpha'}F_{MNK}\partial_\sigma x^K +
\frac{1}{2\pi\alpha'}\mG B_{MN} \right) E^N_{  \
B}\delta(\sigma-\sigma') \ .
\nonumber \\
\end{eqnarray}
With the help of these results we obtain
\begin{eqnarray}\label{pLALambdaGamma}
& &\pb{L^A_\sigma(\Lambda,\sigma),L^B_\sigma(\Gamma,\sigma')}=\nonumber \\
&=&-\frac{1}{T_{D1}}f(\Lambda)f(\Gamma)
(\Gamma+\Lambda)K^{AB}\partial_\sigma\delta(\sigma-\sigma')
-\frac{1}{T_{D1}}K^{AB}\left[\Gamma\frac{df}{d\pi^\sigma}
(\Lambda,\sigma)+\Lambda\frac{df}{d\pi^\sigma}(\Gamma,\sigma)\right]\mG
\delta(\sigma-\sigma')-\nonumber \\
&-&\frac{1}{T_{D1}}f(\Lambda)f(\Gamma)\left(\Lambda+\Gamma
-\frac{1}{T_{D1}}\Lambda\Gamma[(\pi^\sigma+C^{(0)})\kappa+\omega
]\right) K^{AC} f^B_{ \ CE}E^E_{ \ M}\partial_\sigma x^M
\delta(\sigma-\sigma')+\nonumber \\
&+&\frac{\Lambda\Gamma}{T_{D1}^2}f(\Lambda)f(\Gamma) K^{AC}f^B_{ \
CD}K^{DE}E^M_{ \ E}\Pi_M \delta(\sigma-\sigma') \ ,
\nonumber \\
\end{eqnarray}
where we introduced function $f(\Lambda,\sigma)$
\begin{equation}
f(\Lambda,\sigma)=
\frac{1+\Lambda(\pi^\sigma(\sigma)\kappa+\omega)}
{1-\Lambda^2(e^{-2\Phi}+(\pi^\sigma(\sigma)-C^{(0)})^2)}
\end{equation}
and used the fact that
\begin{equation}
\partial_\sigma f(\Lambda,\sigma)=
\frac{df(\Lambda,\sigma)}{d\pi^\sigma}\partial_\sigma \pi^\sigma=
\frac{df(\Lambda,\sigma)}{d\pi^\sigma}\mG \ .
\end{equation}
Now we demand that the expression proportional to $\delta$ function
is equal to
\begin{eqnarray}\label{kLA}
&-&K^{AD}f^B_{ \ DC} (XL^C_\sigma(\Lambda)-YL^C_\sigma(\Gamma))=
-K^{AD}f^B_{ \ DC}(Xf(\Lambda)-Yf(\Gamma))E_M^{ \ C}\partial_\sigma
x^M +\nonumber \\
&+&\frac{1}{T_{D1}}K^{AD}f^B_{ \
DC}(Af(\Lambda)\Lambda-Bf(\Gamma)\Gamma) E^M_{  \ E}K^{CE}\Pi_M \ ,
\nonumber \\
\end{eqnarray}
where $X$ and $Y$ are unknown functions. Comparing
(\ref{pLALambdaGamma}) with (\ref{kLA}) we derive following
equations for $X$ and $Y$
\begin{eqnarray}
& &
\frac{1}{T_{D1}}f(\Lambda)f(\Gamma)(\Lambda+\Gamma-\frac{1}{T_{D1}}
\Lambda \Gamma ((\pi^\sigma+C^{(0)})\kappa
+\omega))=Xf(\Lambda)-Yf(\Gamma) \ , \nonumber \\
& &\frac{\Lambda\Gamma}{T_{D1}}f(\Lambda)f(\Gamma)=X
f(\Lambda)\Lambda-Yf(\Gamma)\Gamma \ . \nonumber \\
\end{eqnarray}
These equations have following solutions
\begin{eqnarray}
X&=&\frac{\Lambda^2}{\Gamma-\Lambda}\frac{f(\Lambda)}{T_{D1}}
[1-\Gamma ((\pi^\sigma+C^{(0)})\kappa+\omega)] \ , \nonumber \\
Y&=&\frac{\Gamma^2}{\Gamma-\Lambda}\frac{f(\Gamma)}{T_{D1}}
[1-\Lambda ((\pi^\sigma+C^{(0)})\kappa+\omega)] \ \nonumber \\
\end{eqnarray}
that is generalization of the solutions found in
\cite{Kluson:2014uaa} to the case of non-trivial Ramond-Ramond flux.
In summary, we obtain final result
\begin{eqnarray}
& &\pb{L^A_\sigma(\Lambda,\sigma),L^B_\sigma(\Gamma,\sigma')}=
\nonumber \\
&=&-\frac{1}{T_{D1}}f(\Lambda)f(\Gamma)
(\Gamma+\Lambda)K^{AB}\partial_\sigma\delta(\sigma-\sigma')
-\frac{1}{T_{D1}}K^{AB}\left(\Gamma\frac{df(\Lambda)}{d\pi^\sigma}
+\Lambda\frac{df(\Gamma)}{d\pi^\sigma}\right)\mG
\delta(\sigma-\sigma')-\nonumber \\
&-&\frac{1}{T_{D1}(\Gamma-\Lambda)} K^{AD}f^B_{ \ DC} \left(
\Gamma^2f(\Gamma) [1-\Lambda ((\pi^\sigma+C^{(0)})\kappa+\omega)]
L^C_\sigma(\Lambda)-\right.\nonumber \\
&-& \left. \Lambda^2f(\Lambda) [1-\Gamma
((\pi^\sigma+C^{(0)})\kappa+\omega)]L^C_\sigma(\Gamma)\right)
\delta(\sigma-\sigma') \ .  \nonumber \\
\end{eqnarray}
We see that the expression proportional to $\mG\approx 0$ vanishes
on the constraint surface. We also see that there is still term
proportional to the derivative of the delta function that needs an
appropriate regularization. Then the terms proportional to the delta
functions are natural generalization of Poisson brackets of flat
connection of principal chiral model with  the Wess-Zumino  term to
the background with RR background two form. Note also that the form
of the expression proportional to the delta functions implies that
corresponding conservative charges are in involution which is the
condition for the integrability of given theory \cite{Dorey:2006mx}.

\section{Explicit Example: D1-Brane on $AdS_3\times S^3$ with
Three Form Fluxes}\label{fourth}
In this section  we will analyze
 D1-brane on $AdS_3\times S^3$ with  three form fluxes. Before we
proceed to the analysis of this specific background we still
consider arbitrary group manifold with non-trivial fluxes but with constant
dilaton and RR zero form. Then  note that with the help of the flat
condition we can rewrite the equation of motion into the form
\begin{equation}\label{conhjA}
\partial_\alpha \hJ^{A\alpha}=0 \ ,
\end{equation}
where we introduced the current
\begin{equation}\label{hJA}
\hJ^{A \alpha}=T_{D1}\left[
\sqrt{e^{-2\Phi_0}+(C^{(0)}-\Pi)^2}\sqrt{-g}g^{\alpha\beta}J^A_\beta+(
\Pi \kappa+\omega) \epsilon^{\alpha\beta}J_\beta^A\right] \ ,
\end{equation}
where $\epsilon^{\tau\sigma}=-\epsilon^{\sigma\tau}=1$.
 We see that the current $\hJ^{A\alpha}$ is conserved.  On the
other hand we see that the current $\hJ^A$ is non-linear and there
is nothing more to say about. We can make given system more
tractable when we introduce an auxiliary metric
$\gamma_{\alpha\beta}$ that obeys the equation
\begin{equation}\label{Tab}
T_{\alpha\beta}\equiv\frac{1}{2}\gamma_{\alpha\beta}
\gamma^{\mu\nu}g_{\mu\nu}-g_{\alpha\beta}=0 \ .
\end{equation}
It is easy to see that this equation has solution
$\gamma_{\alpha\beta}=g_{\alpha\beta}$. If we further introduce
light-cone coordinates
\begin{equation}
\sigma^+=\frac{1}{2}(\tau+\sigma) \ , \quad
\sigma^-=\frac{1}{2}(\tau-\sigma) \
\end{equation}
we can rewrite the equation  (\ref{conhjA}) into the  form
\begin{equation}\label{curconser+-}
\partial_{+}\hJ^{A+}+\partial_{-}\hJ^{A-}=0 \ , \quad
\partial_\pm=\frac{\partial}{\partial \sigma^\pm} \ ,
\end{equation}
where
\begin{eqnarray}
\hJ^{A+}&=&\frac{1}{2}(\hJ^{A\tau}+\hJ^{A\sigma})= \nonumber
\\
&=&\frac{T_{D1}}{2}\left[\sqrt{e^{-2\Phi_0}+(C^{(0)}-\Pi)^2}\sqrt{-\gamma}
\left(\gamma^{\tau\alpha}J_\alpha^A+\gamma^{\sigma\alpha}J_\alpha^A\right)+
(\Pi\kappa+\omega)(J_\sigma^A-J_\tau^A)\right]
 \ ,
\nonumber \\
\hJ^{A-}&=&\frac{1}{2}(\hJ^{A\tau}-\hJ^{A\sigma})= \nonumber \\
&=&\frac{T_{D1}}{2}\left[\sqrt{e^{-2\Phi_0}+(C^{(0)}-\Pi)^2}\sqrt{-\gamma}(\gamma^{\tau
\alpha}J^A_\alpha-\gamma^{\sigma\alpha}J_\alpha^A)+
(\Pi\kappa+\omega)(J_\sigma^A+J_\tau^A)\right] \ . \nonumber \\
\end{eqnarray}
As the next step we fix auxiliary metric to have the form
$\gamma_{\alpha\beta}=\eta_{\alpha\beta} \ ,
\eta_{\mu\nu}=\mathrm{diag}(-1,1)$ keeping in mind that currents
still have to obey the equation (\ref{Tab}). In this gauge
$\hJ^A_\pm$ simplify considerably and we obtain
\begin{eqnarray}\label{hjAgen}
\hJ^{A+}&=& -\frac{1}{2}\hJ^A_-=
\frac{T_{D1}}{2}\left[J_\sigma^A\left(\sqrt{e^{-2\Phi_0}+(C^{(0)}-\Pi)^2}+(
\Pi\kappa+\omega)\right)\right. \nonumber \\
&-& \left. J^A_\tau
\left(\sqrt{e^{-2\Phi_0}+(C^{(0)}-\Pi)^2}+(\Pi\kappa+\omega)\right)\right] \ ,  \nonumber \\
\hJ^{A-}&=&-\frac{1}{2}\hJ^A_+=
-\frac{T_{D1}}{2}\left[J^A_\tau\left(\sqrt{e^{-2\Phi_0}+(C^{(0)}-\Pi)^2}
-(\Pi\kappa+\omega)\right) \right.   \nonumber \\
&+& \left.
J_\sigma^A\left(\sqrt{e^{-2\Phi_0}+(C^{(0)}-\Pi)^2}-(\Pi\kappa+\omega)\right)\right]
\ , \nonumber \\
\end{eqnarray}
where we introduced the light-cone metric with
$\eta_{+-}=\eta_{-+}=-2 \ , \eta^{+-}=\eta^{-+}=-\frac{1}{2}$ so
that $\hJ^{A+}=\eta^{+-}\hJ^A_-=-\frac{1}{2}\hJ_-^A \ ,
\hJ^{A-}=\eta^{-+}\hJ^A_-=-\frac{1}{2}\hJ_+^A$. We see that for
\begin{equation}\label{omegaPicon}
\Pi\kappa+\omega= \sqrt{e^{-2\Phi_0}+(C^{(0)}-\Pi)^2}
\end{equation}
 the current $\hJ^A_+$ vanishes identically and the
equation (\ref{curconser+-}) gives
\begin{equation}\label{con+-}
\partial_{+}\hJ^A_-=0 \ , \quad  \hJ^A_-=2T_{D1}
\sqrt{e^{-2\Phi_0}+(C^{(0)}-\Pi)^2}(J^A_\tau-J^A_\sigma) \ .
\end{equation}
Note that we can write $\hJ_-=\hJ^A_- T_A=2g^{-1}\partial_-g$. Then
from (\ref{con+-}) we obtain
\begin{equation}
\frac{1}{2}\partial_+ \hJ_-=-g^{-1}\partial_+g g^{-1}\partial_-g+
g^{-1}\partial_-\partial_+g
g^{-1}g=g^{-1}\partial_-[\partial_+gg^{-1}]g=0
\end{equation}
so that there is second current $\hJ_+=\partial_+gg^{-1}$ that obeys
the equation
\begin{equation}\label{con-+}
\partial_-\hJ_+=0 \ .
\end{equation}
The equations (\ref{con+-}) and (\ref{con-+}) strongly resembles the
conservations of currents in WZW model.

The previous analysis is valid for any
group manifold with NSNS and RR fluxes and for constant dilaton and
RR zero form. Now we would like to see whether the condition
(\ref{omegaPicon}) can be realized in consistent string background. As the
first case we consider $AdS_3\times S^3\times M$ background with
pure RR flux where $M$ is four torus $T^4$ of four-volume $V_M=(2\pi)^4
v\alpha'^2$ in the metric $ds^2_M$ that implies that each $x^i$ are
identified with the period $2\pi v^{1/4}\alpha'^{1/2}$.
 The background has the form \cite{Maldacena:1998bw}
\begin{eqnarray}\label{D1D5}
ds^2&=&r_1 r_5(ds^2_{AdS_3}+ds^2_{S^3})+\frac{r_1}{r_5}ds^2_M \ , \nonumber \\
F&=&\frac{2r_5^2}{g}(\epsilon+*_6\epsilon_3) \ , \nonumber \\
e^{-\Phi}&=&\frac{1}{g}\frac{r_5}{r_1} \ , \quad
r_5=\sqrt{gQ_5\alpha'} \ , \quad r_1=\frac{4\pi^2
\alpha'}{\sqrt{V_M}}\sqrt{gQ_1\alpha'} \ , \nonumber \\
\end{eqnarray}
where $ds^2_{AdS_3}$ and $ds^2_{S^3}$ are line elements defined with
the group elements  from $SL(2,R)$ and $SU(2)$ respectively that
define the currents $J^A_\alpha$. Further, $ds^2_M$ is a Ricci-flat
metric on $M$ with volume $V_M$ and where $Q_1,Q_5$ are the $D1-$and
$D5-$brane charges. Finally $\epsilon$ is volume element of $AdS_3$
and $*_6\epsilon$ is volume element of $S^3$, where $*_6$ is Hodge
dual in the six dimensions.
Using (\ref{D1D5}) we obtain
 \begin{eqnarray}
\hJ^A_-&=& -\frac{T_{D1}r^2_5}{g}\left[J_\sigma^A\left(\sqrt{1+\Pi^2\frac{g^2
r_1^2}{r_5^2}}+1\right)- J^A_\tau \left(\sqrt{1+\Pi^2\frac{g^2
r_1^2}{r_5^2}}+1\right)\right] \ ,  \nonumber \\
\hJ^A_+&=&\frac{T_{D1}r^2_5}{g}\left[J_\sigma^A\left(\sqrt{1+\Pi^2\frac{g^2
r_1^2}{r_5^2}}-1\right)+ J^A_\tau \left(\sqrt{1+\Pi^2\frac{g^2
r_1^2}{r_5^2}}-1\right)\right] \ .  \nonumber \\
\nonumber \\
\end{eqnarray}
We see that $\hJ^A_+$ vanishes identically in case when $\Pi=0$
while $\hJ^A_-$ is equal to
\begin{equation}
\hJ^A_-= \frac{Q_5}{\pi}(J_\tau^A- J^A_\sigma) \ , \quad  \partial_+\hJ^A_-=0  \ .
  \end{equation}
This is expected result since in this case we have D1-brane in the
near horizon limit of D1-D5-brane system which is S-dual to the
configuration of probe fundamental string in near horizon limit of
the background NS-branes and fundamental strings. These models are
known as WZW models  \cite{Witten:1983ar} and can be analyzed using
powerful conformal field techniques.

Let us now consider D1-brane in this background.  Recall that Type IIB theory
has non-perturbative $SL(2,Z)$ symmetry
\begin{eqnarray}
\hat{G}_{MN}&=&e^{\frac{1}{2}(\hat{\Phi}-\Phi)}G_{MN} \ , \quad
\hat{\tau}=
\frac{a\tau+b}{c\tau+d} \ , \nonumber \\
\hat{B}_{MN}&=&cC^{(2)}_{MN}+d B_{MN} \ , \quad \hat{C}^{(2)}_{MN}
=aC^{(2)}_{MN}+b B_{MN} \
, \nonumber \\
\end{eqnarray}
where $\tau=C^{(0)}+ie^{-\Phi}$ and where $ad-bc=1$. Note that
S-duality transformation corresponds to the following values of
parameters $a=0, b=1,c=-1,d=0$. Then we find that S-dual background
has the form
\begin{eqnarray}\label{Sdual}
e^{-2\hat{\Phi}}&=&\frac{g^2 r^2_1}{r^2_5}=\frac{g^2}{v}\frac{Q_1}{Q_5}  \ , \nonumber \\
d\hat{s}^2&=&e^{-\Phi}ds^2=
\frac{1}{g}r_5^2(ds^2_{AdS_3}+ds^2_{S^3})+gds^2_M=
Q_5\alpha'(ds^2_{AdS_3}+ds^2_{S^3})+g ds^2_M
 \ , \nonumber \\
 H&=&2Q_5\alpha'(\epsilon_3+*_6\epsilon_3) \  \nonumber \\
\end{eqnarray}
so that it is easy to see that the currents $\hJ^A$ have the form
 \begin{eqnarray}
\hJ^A_-&=& -T_{D1}\alpha'Q_5\left[J_\sigma^A\left(\sqrt{\frac{g^2
Q_1}{vQ_5}+\Pi^2}+\Pi\right)- J^A_\tau \left(\sqrt{\frac{g^2
Q_1}{vQ_5}+\Pi^2}+\Pi\right)\right] \ ,  \nonumber \\
\hJ^A_+&=& T_{D1}\alpha'Q_5 \left[J_\sigma^A\left(\sqrt{\frac{g^2
Q_1}{vQ_5}+\Pi^2}-\Pi\right)+ J^A_\tau \left(\sqrt{\frac{g^2
Q_1}{vQ_5}+\Pi^2}-\Pi\right)\right] \ .  \nonumber \\
\nonumber \\
\end{eqnarray}
It is clear that $\hJ^A_+$ does not vanish for finite values
of parameters. On the other hand we easily see
that $\hJ^A_+$ vanishes identically
 when we consider the formal limit
$g\rightarrow 0$. Physically this is the situation when  D1-brane
becomes infinite heavy and decouples so that the probe can be
considered as the collection of $\Pi$ fundamental strings. In this
case the model corresponds to   $\Pi Q_5$ level WZW model that can
 be studied by conventional conformal field theory techniques.
However it is important to stress that this is not possible in case
of finite value of the string coupling constant.

Finally we consider more general case when we perform $SL(2,Z)$
duality transformation from the near horizon limit of D1-D5-brane
background. We begin with the observation \cite{Giveon:1998ns}
that the  near-horizon limit and S-duality commutes. Then for the general form
of $SL(2,Z)$ transformation (with $C^0=0$) we obtain
 (using
$\tau^*=-\tau$)
\begin{eqnarray}\label{Sdualgen}
\hat{C}^{(0)}&=&\frac{ac e^{-2\Phi}+bd}{c^2 e^{-2\Phi}+d^2} \ , \quad
e^{-\hat{\Phi}}=\frac{e^{-\Phi}}{c^2 e^{-2\Phi}+d^2} \ . \nonumber
\\
d\hat{s}^2&=&\sqrt{c^2 e^{-2\Phi}+d^2}ds^2 \ , \quad \hat{B}=c
C^{(2)} \ , \quad \hat{C}^{(2)}=aC^{(2)} \ . \nonumber \\
\end{eqnarray}
Let us start with the symmetric flux background that corresponds following
values of parameters $a,b,c,d$
\begin{equation}
a=1 \ , c=1 \ , b=0 \ , d=1 \ .
\end{equation}
It turns out that in this case  the current
$\hJ^A_+$ vanishes identically in case when $\Pi=0$ while $\hJ^A_-$
is equal to
\begin{equation}
\hJ^A_-= \frac{Q_5}{\pi}(J_\tau^A- J^A_\sigma) \ , \quad
\partial_+\hJ^A_-=0  \ .
\end{equation}
In fact, this remarkable result is valid  whenever the parameter $b$ is
equal to zero. Explicitly, when  $b=0$  we find from the condition
$ad-bc=1$  that
$a=d=1$ that implies
\begin{equation}
\hat{C}^{(0)}=\frac{ce^{-2\Phi}}{c^2 e^{-2\Phi}+d^2} \ , \quad
e^{-\hat{\Phi}}=\frac{e^{-\Phi}}{c^2 e^{-2\Phi}+d^2} \ .
\end{equation}
Then   for $\Pi=0$  and for the background given above we obtain
that the currents (\ref{hjAgen}) have the form
\begin{eqnarray}
\hJ^A_-&=& -\frac{T_{D1}r_5^2}{g}\left(J_\sigma^A - J^A_\tau \right)
\left(\frac{gr_1}{r_5}\sqrt{c^2
e^{-2\Phi}+1}\sqrt{\frac{e^{-2\Phi}}{c^2 e^{-2\Phi}+1}}+1\right)
 \ ,  \nonumber \\
\hJ^A_+&=& \frac{T_{D1}r_5^2}{g}\left(J_\sigma^A + J^A_\tau \right)
\left(\frac{gr_1}{r_5}\sqrt{c^2
e^{-2\Phi}+1}\sqrt{\frac{e^{-2\Phi}}{c^2 e^{-2\Phi}+1}}-1\right) \ ,
\nonumber \\
\end{eqnarray}
where the first square root $ \sqrt{c^2 e^{-2\Phi}+1}$ follows from
the  definition of dual line element (\ref{Sdualgen}) and the second
one from the fact that
$e^{-2\hat{\Phi}}+(C^{(0)})^2=\frac{e^{-2\Phi}}{c^2 e^{-2\Phi}+1}$.
We immediately see that the currents $\hJ^A$ have the same form as
in case of the original near horizon limit of D1-D5-brane background
where $\hJ^A_+$ vanishes identically.

Let us outline results derived in this section. We analyzed the
conditions under which we can find holomorphic or antiholomorphic
currents for D1-brane in the background $AdS_3\times S^3$  with
different combinations of NSNS and RR fluxes. While D1-brane is
integrable for any values of fluxes and world-volume electric flux
it possesses two holomorphic and anti-holomorphic currents that
allow more powerful conformal field theory analysis in near horizon
limit D1-D5-brane background on condition when the electric flux is
zero. We also showed that this hold in case of $AdS_3\times S^3$
with mixed fluxes that is related to the original D1-D5-brane system
by $SL(2,Z)$ duality transformation.

%
%


 \noindent {\bf Acknowledgements}
\\
This work  was supported by the Grant Agency of the Czech Republic
under the grant P201/12/G028.


\begin{thebibliography}{20}




\bibitem{Sfondrini:2014via}
  A.~Sfondrini,
\emph{``Towards integrability for ${\rm Ad}{{{\rm S}}_{{\bf
3}}}/{\rm CF}{{{\rm T}}_{{\bf 2}}}$,''}
  J.\ Phys.\ A {\bf 48} (2015) 2,  023001
  [arXiv:1406.2971 [hep-th]].


\bibitem{Puletti:2010ge}
  V.~G.~M.~Puletti,
\emph{``On string integrability: A Journey through the
two-dimensional hidden symmetries in the AdS/CFT dualities,''}
  Adv.\ High Energy Phys.\  {\bf 2010} (2010) 471238
  [arXiv:1006.3494 [hep-th]].

\bibitem{vanTongeren:2013gva}
  S.~J.~van Tongeren,
\emph{``Integrability of the ${\rm Ad}{{{\rm S}}_{5}}\times {{{\rm
S}}^{5}}$ superstring and its deformations,''}
  J.\ Phys.\ A {\bf 47} (2014) 433001
  [arXiv:1310.4854 [hep-th]].



\bibitem{Beisert:2010jr}
  N.~Beisert {\it et al.},
\emph{``Review of AdS/CFT Integrability: An Overview,''}
  Lett.\ Math.\ Phys.\  {\bf 99} (2012) 3
  [arXiv:1012.3982 [hep-th]].

\bibitem{Bena:2003wd}
  I.~Bena, J.~Polchinski and R.~Roiban,
\emph{``Hidden symmetries of the AdS(5) x S**5 superstring,''}
  Phys.\ Rev.\ D {\bf 69} (2004) 046002
  [hep-th/0305116].
\bibitem{Dorey:2006mx}
  N.~Dorey and B.~Vicedo,
\emph{``A Symplectic Structure for String Theory on Integrable
Backgrounds,''}
  JHEP {\bf 0703} (2007) 045
  [hep-th/0606287].




\bibitem{Giveon:1998ns}
  A.~Giveon, D.~Kutasov and N.~Seiberg,
\emph{``Comments on string theory on AdS(3),''}
  Adv.\ Theor.\ Math.\ Phys.\  {\bf 2} (1998) 733
  [hep-th/9806194].

\bibitem{Elitzur:1998mm}
  S.~Elitzur, O.~Feinerman, A.~Giveon and D.~Tsabar,
\emph{``String theory on $AdS(3) x S**3 x S**3 x S**1$,''}
  Phys.\ Lett.\ B {\bf 449} (1999) 180
  [hep-th/9811245].

\bibitem{deBoer:1998pp}
  J.~de Boer, H.~Ooguri, H.~Robins and J.~Tannenhauser,
\emph{``String theory on AdS(3),''}
  JHEP {\bf 9812} (1998) 026
  [hep-th/9812046].

\bibitem{Maldacena:2000hw}
  J.~M.~Maldacena and H.~Ooguri,
\emph{``Strings in AdS(3) and SL(2,R) WZW model 1.: The Spectrum,''}
  J.\ Math.\ Phys.\  {\bf 42} (2001) 2929
  [hep-th/0001053].

\bibitem{Maldacena:2000kv}
  J.~M.~Maldacena, H.~Ooguri and J.~Son,
\emph{``Strings in AdS(3) and the SL(2,R) WZW model. Part 2.
Euclidean black hole,''}
  J.\ Math.\ Phys.\  {\bf 42} (2001) 2961
  [hep-th/0005183].

\bibitem{Maldacena:2001km}
  J.~M.~Maldacena and H.~Ooguri,
\emph{``Strings in AdS(3) and the SL(2,R) WZW model. Part 3.
Correlation functions,''}
  Phys.\ Rev.\ D {\bf 65} (2002) 106006
  [hep-th/0111180].

\bibitem{Berkovits:1999im}
  N.~Berkovits, C.~Vafa and E.~Witten,
\emph{``Conformal field theory of AdS background with Ramond-Ramond
flux,''}
  JHEP {\bf 9903} (1999) 018
  [hep-th/9902098].


\bibitem{Chen:2005uj}
  B.~Chen, Y.~L.~He, P.~Zhang and X.~C.~Song,
\emph{``Flat currents of the Green-Schwarz superstrings in AdS(5) x
S**1 and AdS(3) x S**3 backgrounds,''}
  Phys.\ Rev.\ D {\bf 71} (2005) 086007
  [hep-th/0503089].

\bibitem{Babichenko:2009dk}
  A.~Babichenko, B.~Stefanski, Jr. and K.~Zarembo,
\emph{``Integrability and the AdS(3)/CFT(2) correspondence,''}
  JHEP {\bf 1003} (2010) 058
  [arXiv:0912.1723 [hep-th]].

\bibitem{Quella:2007sg}
  T.~Quella, V.~Schomerus and T.~Creutzig,
\emph{``Boundary Spectra in Superspace Sigma-Models,''}
  JHEP {\bf 0810} (2008) 024
  [arXiv:0712.3549 [hep-th]].

\bibitem{Cagnazzo:2012se}
  A.~Cagnazzo and K.~Zarembo,
\emph{``B-field in AdS(3)/CFT(2) Correspondence and
Integrability,''}
  JHEP {\bf 1211} (2012) 133
   [JHEP {\bf 1304} (2013) 003]
  [arXiv:1209.4049 [hep-th]].



\bibitem{Hoare:2013pma}
  B.~Hoare and A.~A.~Tseytlin,
\emph{``On string theory on $AdS_3 x S^3 x T^4$ with mixed 3-form
flux: tree-level S-matrix,''}
  Nucl.\ Phys.\ B {\bf 873} (2013) 682
  [arXiv:1303.1037 [hep-th]].
\bibitem{Hoare:2013ida}
  B.~Hoare and A.~A.~Tseytlin,
\emph{``Massive S-matrix of $AdS_3 x S^3 x T^4$ superstring theory
with mixed 3-form flux,''}
  Nucl.\ Phys.\ B {\bf 873} (2013) 395
  [arXiv:1304.4099 [hep-th]].

\bibitem{Hoare:2013lja}
  B.~Hoare, A.~Stepanchuk and A.~A.~Tseytlin,
\emph{``Giant magnon solution and dispersion relation in string
theory in $AdS_3$x$S^3$x$T^4$ with mixed flux,''}
  Nucl.\ Phys.\ B {\bf 879} (2014) 318
  [arXiv:1311.1794 [hep-th]].

\bibitem{Babichenko:2014yaa}
  A.~Babichenko, A.~Dekel and O.~Ohlsson Sax,
\emph{``Finite-gap equations for strings on AdS$_{3}$ x S$^{3}$ x
T$^{4}$ with mixed 3-form flux,''}
  JHEP {\bf 1411} (2014) 122
  [arXiv:1405.6087 [hep-th]].



\bibitem{Lloyd:2014bsa}
  T.~Lloyd, O.~Ohlsson Sax, A.~Sfondrini and B.~Stefañski, Jr.,
\emph{``The complete worldsheet S matrix of superstrings on $AdS_3 x
S^3 x T^4$ with mixed three-form flux,''}
  Nucl.\ Phys.\ B {\bf 891} (2015) 570
  [arXiv:1410.0866 [hep-th]].

\bibitem{Sundin:2014ema}
  P.~Sundin and L.~Wulff,
\emph{``One- and two-loop checks for the $AdS_3$ x $S^3$ x $T^4$
superstring with mixed flux,''}
  J.\ Phys.\ A {\bf 48} (2015) 10,  105402
  [arXiv:1411.4662 [hep-th]].




\bibitem{Kluson:2014uaa}
  J.~Kluson,
\emph{``Integrability of D1-brane on Group Manifold,''}
  JHEP {\bf 1409} (2014) 159
  [arXiv:1407.7665 [hep-th]].



\bibitem{Maillet:1985ek}
  J.~M.~Maillet,
\emph{``New Integrable Canonical Structures in Two-dimensional
Models,''}
  Nucl.\ Phys.\ B {\bf 269} (1986) 54.

\bibitem{Delduc:2012qb}
  F.~Delduc, M.~Magro and B.~Vicedo,
\emph{``Alleviating the non-ultralocality of coset sigma models
through a generalized Faddeev-Reshetikhin procedure,''}
  JHEP {\bf 1208} (2012) 019
  [arXiv:1204.0766 [hep-th]].

\bibitem{Delduc:2012vq}
  F.~Delduc, M.~Magro and B.~Vicedo,
\emph{``Alleviating the non-ultralocality of the $AdS_5 x S^5$
superstring,''}
  JHEP {\bf 1210} (2012) 061
  [arXiv:1206.6050 [hep-th]].

\bibitem{Evans:2000hx}
  J.~M.~Evans, M.~Hassan, N.~J.~MacKay and A.~J.~Mountain,
\emph{``Conserved charges
 and supersymmetry in principal chiral and WZW models,''}
  Nucl.\ Phys.\ B {\bf 580} (2000) 605
  doi:10.1016/S0550-3213(00)00257-1
  [hep-th/0001222].

\bibitem{Witten:1983ar}
  E.~Witten,
\emph{``Nonabelian Bosonization in Two-Dimensions,''}
  Commun.\ Math.\ Phys.\  {\bf 92} (1984) 455.
  doi:10.1007/BF01215276

\bibitem{Maldacena:1998bw}
  J.~M.~Maldacena and A.~Strominger,
\emph{``AdS(3) black holes and a stringy exclusion principle,''}
  JHEP {\bf 9812} (1998) 005
  doi:10.1088/1126-6708/1998/12/005
  [hep-th/9804085].

\bibitem{Giveon:1998ns}
  A.~Giveon, D.~Kutasov and N.~Seiberg,
\emph{``Comments on string theory on AdS(3),''}
  Adv.\ Theor.\ Math.\ Phys.\  {\bf 2} (1998) 733
  [hep-th/9806194].


\end{thebibliography}
\end{document}